\documentclass[
superscriptaddress,
showpacs,preprintnumbers,
nobibnotes,
amsmath,amssymb,
aps,
prl,
twocolumn,
floatfix,
]{revtex4-1}

\pdfoutput=1

\usepackage{graphicx}
\usepackage{dcolumn}
\usepackage{bm}
\usepackage{color}
\usepackage{comment}

\usepackage{ulem}

\usepackage{xurl}
\usepackage{hyperref}
\usepackage{natbib}
\normalsize
\usepackage{booktabs} 

\begin{document}

\title{Complex topological features of reservoirs shape learning performances in bio-inspired recurrent neural networks}

\author{Valeria d'Andrea}
\email[Corresponding author:~]{vdandrea@fbk.eu}%
\affiliation{Fondazione Bruno Kessler, Via Sommarive 18, 38123 Povo (TN), Italy}

\author{Michele Puppin}
\affiliation{Dipartimento di Fisica e Astronomia "Galileo Galilei", Università di Padova, Via Marzolo 8, 35131 Padova, Italy}

\author{Manlio De Domenico}
\email[Corresponding author:~]{manlio.dedomenico@unipd.it}%
\affiliation{Dipartimento di Fisica e Astronomia "Galileo Galilei", Università di Padova, Via Marzolo 8, 35131 Padova, Italy}

\date{\today}


\begin{abstract}
Recurrent networks are a special class of artificial neural systems that use their internal states to perform computing tasks for machine learning. One of its state-of-the-art developments, i.e. reservoir computing (RC), uses the internal structure -- usually a static network with random structure -- to map an input signal into a nonlinear dynamical system defined in a higher dimensional space. Reservoirs are characterized by nonlinear interactions among their units and their ability to store information through recurrent loops, allowing to train artificial systems to learn task-specific dynamics. However, it is fundamentally unknown how the random topology of the reservoir affects the learning performance.
Here, we fill this gap by considering a battery of synthetic networks -- characterized by different topological features -- and 45 empirical connectomes -- sampled from brain regions of organisms belonging to 8 different species -- to build the reservoir and testing the learning performance against a prediction task with a variety of complex input signals.  We find nontrivial correlations between RC performances and both the number of nodes and rank of the covariance matrix of activation states, with performance depending on the nature -- stochastic or deterministic -- of input signals. Remarkably, the modularity and the link density of the reservoir are found to affect RC performances: these results cannot be predicted by models only accounting for simple topological features of the reservoir. Overall, our findings highlight that the complex topological features characterizing biophysical computing systems such as connectomes can be used to design efficient bio-inspired artificial neural networks.
\end{abstract}

\maketitle

\section*{\label{sec:intro}Introduction}
The physical and biological basis of cognition are embedded in the complex temporally and spatially multiscale structure of the brain~\cite{bullmore2012economy,Bassett2011,de2014graph,presigny2022colloquium}.
Even if it is not yet fully understood how cognitive functions can emerge from brain activity, it has been demonstrated that structural properties of human brain networks play a key role in brain functioning. For instance, in healthy individuals the configuration of brain connectivity is built upon different aspects, such as behavioral variability \cite{Bassett2009Cognitive}, cognitive ability \cite{Li2009,vandenHeuvel2009},  age \cite{Meunier2008,Sifis2009} and gender \cite{Gong2009}. Connectivity patterns exhibit plasticity: they are inherently dynamical and reshaped as a consequence of experimental tasks \cite{Bassett2007,DeVicoFallani2008} and drug treatment \cite{Achard2007,Schwarz2009}. Similarly, individuals affected by psychiatric and neurodegenerative disorders present abnormal structural and functional connectivity between brain regions \cite{Catani2005}. Some diseases characterized by abnormal cortical structural/functional organization are schizophrenia \cite{Lynall2010,de2016mapping},  Alzheimer’s disease \cite{Stam2008,guillon2017loss}, epilepsy \cite{Horstmann2010} and many others.

The characterization of the human brain architecture requires also investigation from a dynamical perspective \cite{Deco2011,Churchland2010}, which has been successfully investigated with methods from statistical physics~\cite{Burioni2014-dp,Di_Santo2018-gx,munoz2018colloquium}. In the quest to build computational models resembling the most salient feature of the human brain, Reservoir Computing (RC)~\cite{Lukosevicius2009,gauthier2021next} provides a suitable framework for its transparency and its versatility for implementation on physical devices~\cite{paquot2012optoelectronic,van2017advances,tanaka2019recent}. RC identifies a class of algorithms that are biologically-inspired computing systems composed by a set of processing units called neurons, organized within a network. A neuron receives a signal, processes it and propagates the output signal to the connected neurons. The strength of the connection, analogously to the synaptic strength, encodes how much a neuron influences another neuron. By adapting network structure, i.e. finding the optimal set of link weights, the machine can learn to perform desired tasks and to develop meaningful internal representations of the environment. RC algorithms represent the state-of-the-art algorithms to tackle tasks such as images recognition, text comprehension, natural language processing, game playing and many others. In these tasks, RC algorithms have reached super-human performances \cite{Storrs2019,Pushparaja2018}.

From a network neuroscience point of view, RC-based approaches can be used to gain new insights about the role of brain connectivity patterns in the execution of complex cognitive tasks. In fact, at variance with other Machine Learning methods, RC approaches are characterized by biological plausibility, with several architectural and dynamical properties of mammalian brains used in RC design \cite{Lukosevicius2009}. RC can provide insights in how brains perform accurate computations with an inaccurate physical scaffolding \cite{Buonomano1995,Hausler2007} and it helps explaining brain time encoding \cite{Karmarkar2007}, visual information processing in primary visual cortex \cite{Stanley1999,Nikolic2006} and the representation of sequential information. In addition, RC offers a functional interpretation of the cerebellar circuitry \cite{Kistler2002,Yamazaki2007}. 
The structure of RC networks plays a fundamental role, allowing to approximate very complex functions and it has been shown that real-world inspired connectivity such as Watts-Strogatz and Barabási-Albert models has resulted in competitive performances compared to optimized state-of-the-art architectures \cite{Xie2019}. Furthermore, during training feedforward networks have proved to spontaneously form non-random topologies also found in brains, such as modular structures \cite{Filan2020}. This and other evidences support the notion that non-random topologies can lead to desired performance of ANNs \cite{Damicelli2021}. 

In our work we probe the inverse approach by using real-world connectomes as reservoirs, in order to investigate how RC performance in a prediction task is affected by topological features occurring in empirical network structures. We also investigate the dynamical input representation by studying the dimensionality of elicited connectome internal states.

To verify if the RC behavior is affected by the topological features identified in real-world connectomes, we test our results on three null models of increasing complexity, namely an Erd\H{o}s R\'{e}nyi (ER) model, Configuration Model (CM) and a Stochastic Block Model (SBM).

Finally, we test our results on multiple signals to verify if the RC performance depends on the stochastic or deterministic nature of the input signal.

\section*{\label{sec:results}Results}
In this study, we collected and use 45 connectomes providing a network representation of the neuronal system of C. elegans, C. intestinalis, Drosophila, Human, Mouse, Platynereis dumerilly, Rattus norvegicus and Rhesus macaque in terms of adjacency matrices $A_{conn}$. For each connectome, we use the corresponding adjacency matrix to build a reservoir.

\begin{figure}
\centering
\includegraphics[width=\linewidth]{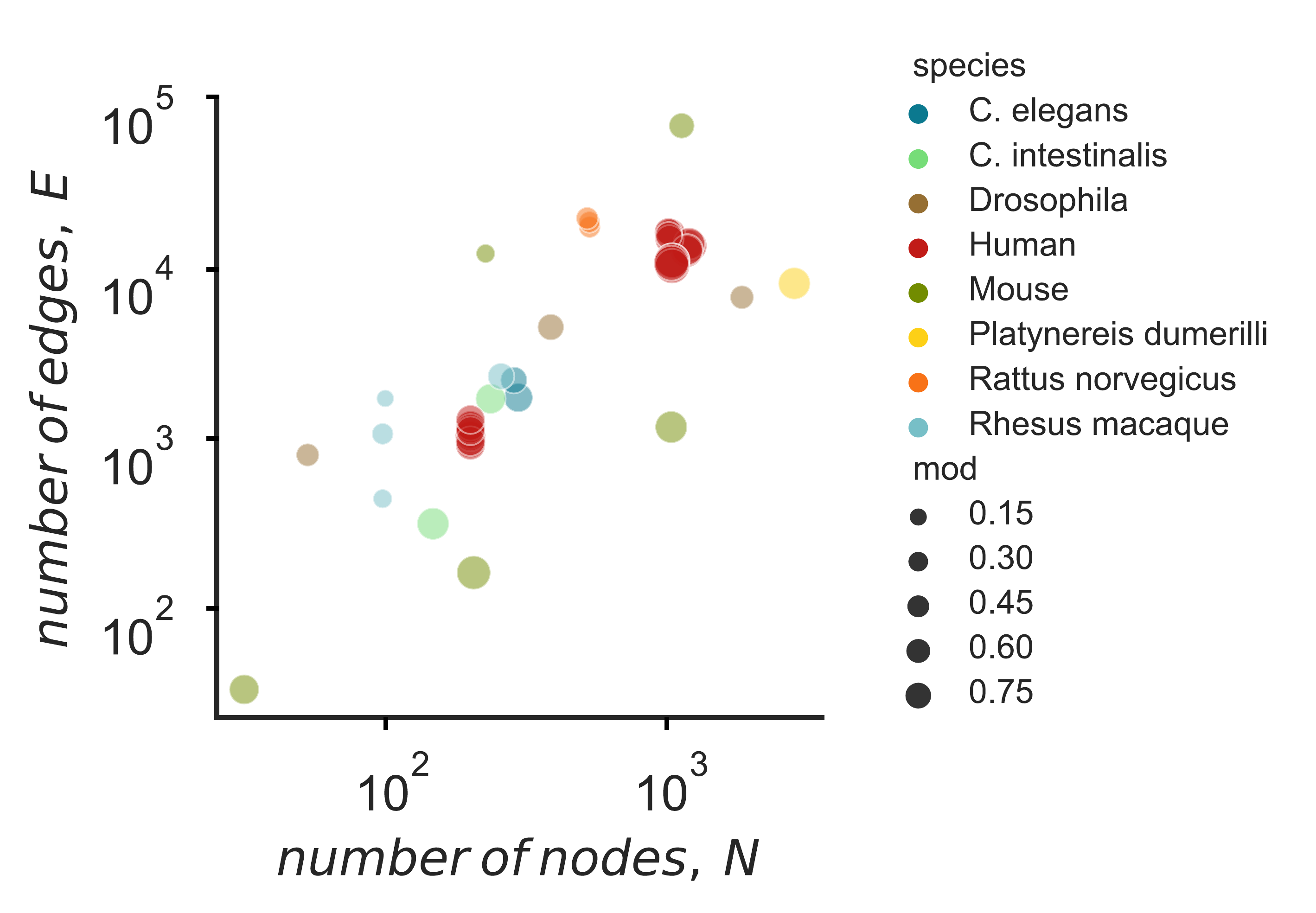}
\caption{\textbf{Allometric scaling in empirical connectomes.} Scatter plot describing the number of neuronal links, $E$, as a function of the number of nodes, $N$, for the connectome data sets ($n=45$ networks) considered in this study. Each point is a single network, its color identifies the species and its size is proportional to the network modularity (mod). }
\label{fig:figure1}
\end{figure}

For each empirical network we calculate some topological indicators, such as  average degree and modularity, which are widely used to characterize network structures and their mesoscale organization. Figure~\ref{fig:figure1} shows that connectome data span more than 2 orders of magnitude in the number of nodes and more than 3 orders of magnitude in number of edges; a complete list of connectomes is shown in Tab. \ref{tab:info_connect} along with the number of nodes and the average degree of the network.

The performances of the connectome-based reservoirs are tested in a one-step-ahead prediction task, where given a point of a temporal sequence as input the model is asked to predict the next point of the sequence. To probe the ability of our bio-inspired machines to deal with signals of different nature, we consider four different time series as inputs: two stochastic signals (white noise and fractional Brownian noise) and two deterministic chaotic signals (Ikeda and Mackey-Glass signals). 

\begin{table*}[]
\centering
\begin{tabular}{lcccc}
\toprule
\textbf{Organism} & \textbf{Num. of} & \textbf{Num. of} & \textbf{Brain mass} &  \textbf{Brain volume} \\
 & \textbf{neurons} & \textbf{synapses} & \\ 
&&& $(\mathbf{\text{g}})$ &  $(\mathbf{\text{cm}^3})$ \\
\midrule
C. intestinalis & $213$ & $3105$ & - & - \\
C. elegans & $302$ & $7500$ & - & 1 \\
Platynereis dumerilli & $2728$ & $1.14\cdot10^4$ & - & - \\
Drosophila & $10^5$ & $10^7$ & $3.5\cdot10^{-5}$ & $0.08$ \\
Mouse & $7\cdot10^7$ & $10^{12}$ & $0.36$ & $440$ \\
Rattus norvegicus & $2\cdot10^8$ & $5\cdot10^{11}$ & $1.8$ & $1200$ \\
Rhesus & $6\cdot10^9$ & - & $75$ & $94968$ \\
Human & $10^{11}$ & $10^{14}$ & $1300$ & $1.2\cdot10^6$ \\
\bottomrule
\end{tabular}
\caption{\textbf{Neurophysiology of the empirical connectomes considered in this work.} Information about the number of neurons, number of synapses, brain mass (in grams) and brain volume (in cm$^3$) for C. elegans, C. intestinalis, Drosophila, Human, Mouse, Platynereis dumerilly, Rattus norvegicus and Rhesus macaque.}
\label{tab:species_info}
\end{table*}

\begin{figure*}
\centering
\includegraphics[width=\linewidth]{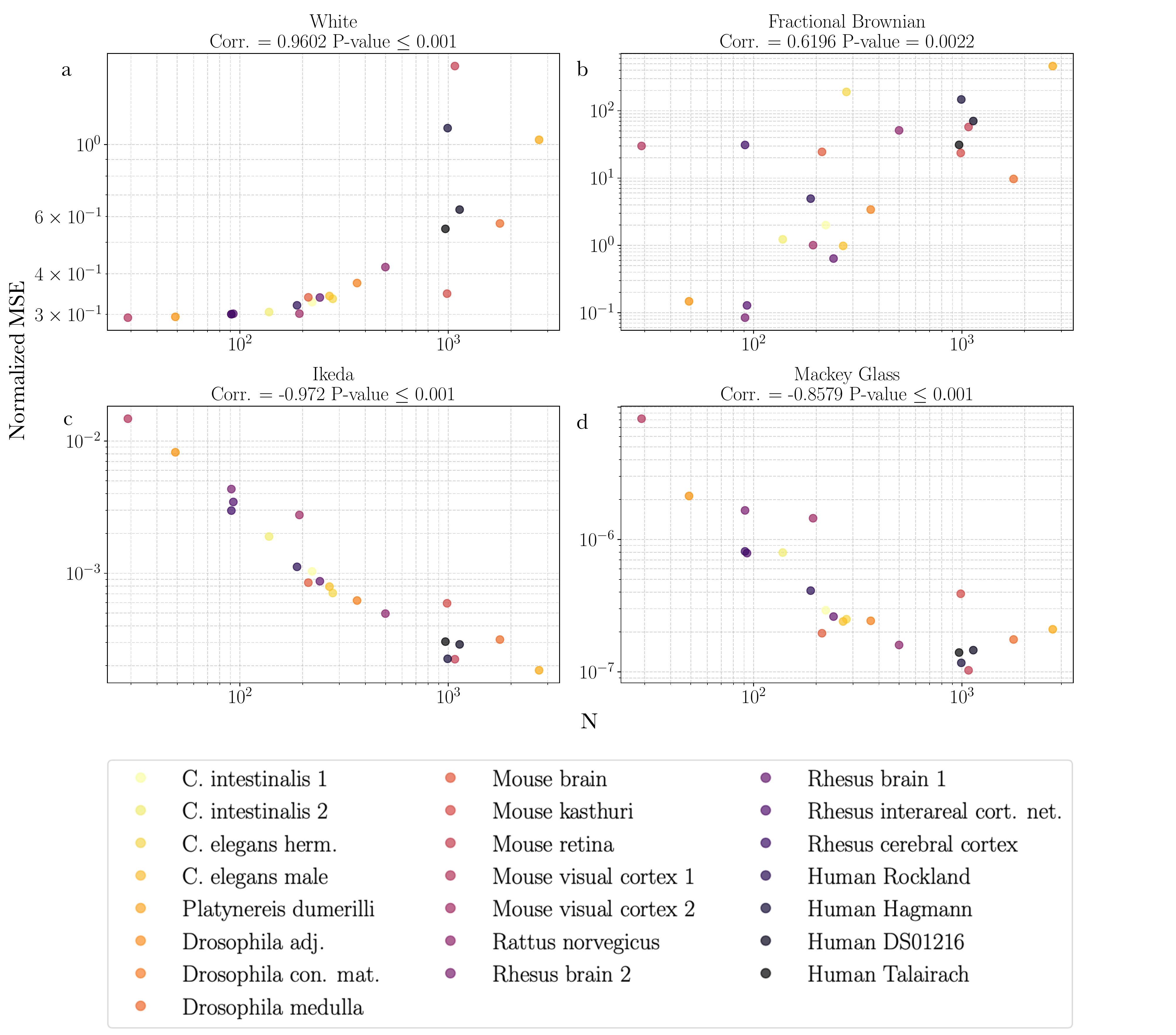}
\caption{\textbf{Prediction errors depend on system's size and the nature of input signals.} Correlations between normalized MSE and number of nodes $N$. a) white, b) fractional Brownian, c) Ikeda and d) Mackey Glass. Each point represents the average across all networks of the same organism with the same number of nodes. The color darkness reflects the total number of neurons in the organism's brain (Tab.~\ref{tab:species_info}). Correlations are significant ($p-value \leq 0.05$) for all signals. Positive values are measured for white and fractional Brownian signals, while negative values are measured for Ikeda and Mackey Glass signals.
}
\label{fig:figure2}
\end{figure*}

For each combination of connectome and input signal, the corresponding reservoir is trained and the activation states are stored in matrix $X$. Then we compute the rank of the covariance matrix of $X$, as it measures the number of independent (uncorrelated) vector states and can be used to evaluate the dimensionality of the reservoir as a dynamical system \cite{Carrol2019}. Consequently, we evaluate how RC performances -- here measures as normalized MSE -- vary as a function of different network measures, namely number of nodes $N$, average degree $\langle k\rangle$, rank $r$ of the covariance matrix of activation states and network modularity $Q$.
Correlations between performances and networks' macroscopic properties are quantified by the Spearman coefficient, whose p-value is estimated using a permutation test, that is by resampling without replacements $N$ and the normalized MSE, and then evaluating the Spearman correlation for the resampled data. This procedure is independently repeated $n=10,000$ times. 

Figure~\ref{fig:figure2} shows RC performance as a function of the number of nodes $N$ of the connectomes. On the one hand, results show that the RC is actually able to learn the chaotic dynamics of Ikeda and Mackey-Glass deterministic signals and to reproduce them over time with a good approximation, with errors that range between $10^{-2}$ and $10^{-7}$ (arbitrary units). On the other hand, the performance of RC is lower against white and fractional Brownian noisy signals, where prediction errors range between $10^{-1}$ and $10^{2}$ as expected due to the highly stochastic nature of those sequences.

Correlations are found to be all significant; positive values are observed for white and fractional Brownian signals while negative values are observed for Ikeda and Mackey-Glass signals. Thus, for stochastic signals, the error increases as the number of nodes increases, while for deterministic chaotic signals the error decreases as the number of nodes increases. No correlations are found between RC performance and the total number of neurons in the organism brain from which the connectome is obtained (represented in different colors in Fig. \ref{fig:figure2}, see also Table \ref{tab:species_info}). Similar correlation values and statistical significances are observed for the average degree (Supplementary Fig. S1), the rank of the covariance matrix of activation states (Supplementary Fig. S2) and the network modularity (Supplementary Fig. S3).

Then we test the hypothesis that the observed results are consistent with the topology of a set of synthetic data generated with three different generative models preserving one or more features of the original connectomes. With this approach, we probe which topological features characterizing empirical connectomes are mostly responsible for the measured performances of the machine. In this work, we consider three different classes of null models: an Erd\H{o}s R\'{e}nyi (ER) random network, that only preserves the number of nodes and the number of edges of the real connectome; a Configuration Model (CM), that reproduces the degree distribution but destroys degree-degree correlations and other structures observed in the real networks; and a Stochastic Block Model (SBM), where the block structure is the topological feature that is preserved by the synthetic network. 

For each connectome, 20 independent realizations of each generative model are considered, and for each realization the RC algorithm is run for one random initialization of the weights and the average performance over all the realizations is computed. To compare empirical and synthetic measures we use a Z-score: negative scores indicate that the measure computed with the empirical network is significantly greater than the one of the corresponding synthetic network; conversely, positive scores indicate the opposite behavior. The two measures we compare in empirical and synthetic networks include the rank of the covariance matrix of activation states and the normalized MSE.

\begin{figure*}
\centering
\includegraphics[width=\linewidth]{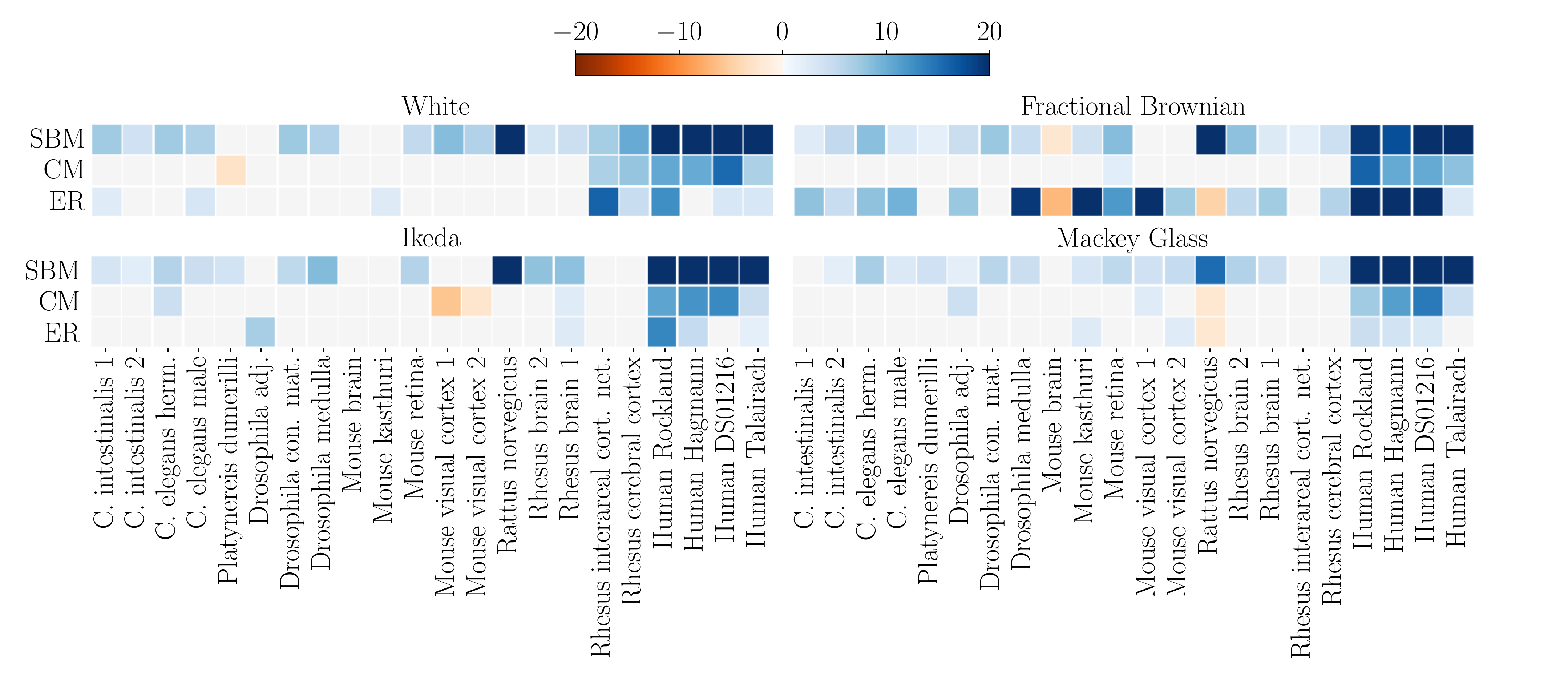}
\caption{\textbf{Number of independent activation states increases with null model complexity and organism brain volume.} Heatmap of Z-scores between the rank of the covariance matrix of activation states obtained in empirical networks and in Erd\H{o}s R\'{e}nyi (ER), Configuration Model (CM) and Stochastic Block Model (SBM) networks. Positive Z-scores (blue) indicate that the number of independent nodes of the empirical network is significantly lower than the one of the corresponding synthetic network; negative Z-scores (brown) indicate that the number of independent vector states is significantly greater in real connectomes. Light gray indicates non significant results (Z-score in $[-2,2]$). Positive Z-scores prevail for all signals and, across different synthetic networks, higher absolute values of the Z-score are found as the complexity of the null model increases. High absolute values of the Z-score are prevalent for Human connectomes.}
\label{fig:figure3}
\end{figure*}

The number of independent activation states of the network, here measured as the rank of their covariance matrix, characterizes the dimension of the network actually involved in the requested task, as it discounts all the nodes that follow identical trajectories. Higher ranks have been found to correlate with lower error prediction \cite{Carrol2019, Lukosevicius2009}. Figure~\ref{fig:figure3} shows the heatmaps representing the differences in the rank of the covariance matrix between empirical networks and the corresponding synthetic networks. Positive Z-scores prevail regardless of the nature of the input signal. At fixed connectome, higher values are found as the complexity of the null model increases, that is, the highest rank is found in SBM model while the lowest is found in ER model. This result suggests that, among the network measures that characterize an empirical connectome, the block structure is the one that mainly could affect prediction performance.
Along the same synthetic model, higher values of the Z-score are prevalent for human connectomes and, in general, the number of independent nodes increases with the number of neurons of the considered organism (see Table \ref{tab:species_info}). Our results attest that synthetic networks, and in particular SBM, have a higher number of independent vector states involved in the input prediction task and this would suggest lower prediction errors for all the considered signals. 

\begin{figure*}
\centering
\includegraphics[width=\linewidth]{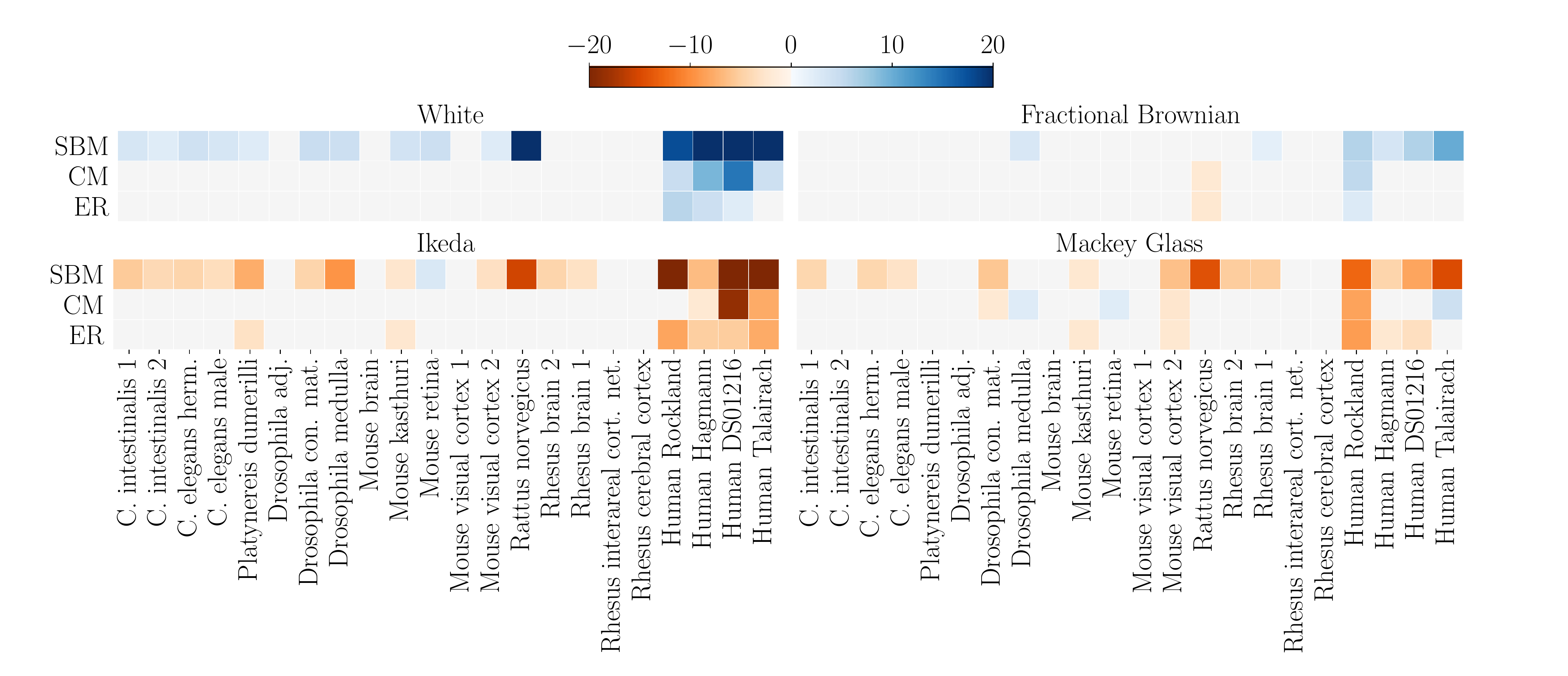}
\caption{\textbf{Topological features and input dynamics affect prediction performance.} Heatmap of Z-scores between normalized MSE obtained in empirical networks and normalized MSE in  Erd\H{o}s R\'{e}nyi (ER), Configuration Model (CM) and Stochastic Block Model (SBM) networks. Positive Z-score (blue) indicate that the performances of the empirical network are significantly better than the ones of the corresponding synthetic network; negative Z-scores (brown) indicate that the performances of the synthetic network are significantly better than the ones of the empirical network. Light gray indicates non significant results (Z-score in $[-2,2]$). Positive Z-scores prevail for white and fractional Brownian signals. Negative values, instead, prevail for Ikeda and Mackey-Glass signals. High absolute values of the Z-score are prevalent for Human connectomes. Also, higher absolute values of the Z-score are found as the complexity of the null model increases.}
\label{fig:figure4}
\end{figure*}

To test this hypothesis, and to verify if the trends of prediction performances we found in empirical connectomes actually depend on topological features, we compare errors in RC performances between empirical and synthetic connectomes. Results are shown in Fig. \ref{fig:figure4} and indicate that, when the input signal is deterministic, as for Ikeda and Mackey Glass signals, as expected the normalized MSE is greater for real connectome reservoirs with respect to the case in which the reservoir is a synthetic model. For white and fractional Brownian input signals, we find that the performances of the empirical connectomes are significantly better than the ones of the synthetic null models. These results could be explained by overfitting occurring, for stochastic signals, when the number of independent network nodes increases.

\section{Discussion}
In the present work, we exploit the framework of RC to investigate the relationship between structure and function in real brain connectomes. In practice, we use that state-of-the-art machine learning technique to gain insights about empirical neural systems in terms of how the underlying network structure can influence the performances of a prediction task.

Previous studies used biologically-inspired structures to perform computational tasks and showed that the network topology could relevantly affect neural computation and achieved valuable performance in a requested task \cite{Bray1995, Damicelli2021,Suarez2021}. Our work capitalizes on those findings and clarifies which topological features, typical of real networks, are relevant for computation. To this purpose, we used a wide data set of 45 connectomes extracted from 8 different species, validating our results by means of different null synthetic models with increasing topological complexity. We found that multiple features correlate with RC performance and that, in particular, a modular structure has a significant impact on prediction error.
This result is in agreement with studies that suggested how functional \cite{Meunier2009} and structural \cite{Chen2008,Bassett2010} hierarchical modularity \cite{Meunier2009,Werner2010} is characteristic of human brain connectivity.  Block structure produces a compartmentalization that reduces the interdependence of modules and enhances robustness \cite{Kirschner1998,Felix2008}
and allows the formation of complex hierarchical architectures that promote a high functional specificity of subsystems \cite{Bassett2011}.
Furthermore, brain modular structure allows different regions to act either as hubs of high connectivity or as peripheral nodes of local processing \cite{Bassett2011,Rodriguez2019}. 
In addition to robustness enhancement and specificity, modularity facilitates behavioral adaptation \cite{Kirschner1998} because each subsystem can modify its function without altering other subsystems functioning. 
The fact that real connectomes achieve a lower performance than the corresponding modular synthetic model can be explained by considering the wiring cost that real brain networks have to minimize: synapses functioning contributes to most of the energy consumption in the human brain \cite{Lennie2003} thus hierarchical modularity of brain structure needs to be compatible with an evolutionary pressure to minimize energy consumption while building and maintaining brain connectivity \cite{Bassett2010,Niven2008}. Short-range connections between neurons are predominant in real brain networks \cite{Harvat2016}, while wiring cost constraints are not affecting synthetic generative models. 

Another measure that we find to affect RC performances is the dimensionality of the reservoir as a dynamical system, that is the number of independent vector states. 
High reservoir dimensionality is fundamental to ensure the input representation is rich enough for the readout training procedure to work properly. It helps the separation of initially inseparable inputs when performing classification tasks \cite{Tanaka2019}. Moreover, it allows learning spatio-temporal dependencies of inputs when performing prediction tasks \cite{Tanaka2019}. Usually, reservoirs contain a large number of neurons with order ranging from tens to thousands \cite{Lukosevicius2009}. An interesting result of our work is that differences in ranks between empirical and synthetic networks depend on the number of neurons of the considered organisms, suggesting that wiring patterns change as a function of the species' complexity.

Moreover, we run our analysis for different input signals and we find that, unlike what we could expect from the number of independent vector states, when the signals that have to be predicted are noisy the performances for reservoirs with modular structures are lower than the one evaluated with empirical connectomes. 
Our finding could be computationally explained by an overfitting effect and is in agreement with the results of previous works, according to which biological networks have optimized topologies capable of suppressing noisy fluctuations that consist in sparser and more hierarchically organized structures \cite{Ronellenfitsch2018}.

There are also some limitations of our study that have to be considered in the interpretation of the results. First, we study the reservoir dynamics at a fixed amount of memory and nonlinearity, while it has been shown that performances could vary depending on the underlying dynamical regime \cite{Suarez2021}.
Second, we collected our connectome data set from several studies. Even though this has the advantage to offer an exhaustive analysis across several organisms and species, the different experimental methodologies that have been used to reconstruct connectomes could affect our comparisons. 
Finally, in this work, we focus on a specific task, namely the prediction of a signal provided in input to the reservoir, that represents a narrow window on the complexity of the cognitive functions performed by the brain. In order to better understand how cognition and perception are implemented in the brain, other processes should be investigated, including memory and classification as well as dynamic regimes with critical transitions that are typical of the brain.

\section*{Material and Methods}
\begin{small}

\subsection*{Connectome and synthetic networks data}

For each empirical connectome analogous synthetic networks are generated as null models of increasing complexity to verify if the behavior of the empirical networks actually depends on topological correlations. Erd\H{o}s R\'{e}nyi (ER) model, Configuration Model (CM) and Stochastic Block Model (SBM) are used to generate random graphs with the same number of nodes of the corresponding empirical network. With ER, CM and SBM it is possible to produce synthetic networks that randomize the network topology while preserving respectively:
\begin{itemize}
    \item[] (ER) the average degree,
    \item[] (CM) the degree distribution,
    \item[] (SBM) the degree distribution and the block structure
\end{itemize}
of the empirical network.

For each connectome (45) and signal (8) the algorithm is run with 20 different random initializations of the weights and the average performances over all the realizations are computed.
Results referring to networks of the same organism with the same number of neurons are averaged as well if convenient.

\begin{table}[ht]
\small
\centering
\begin{tabular}{lcccc}
\toprule
\textbf{Network label} & \textbf{Num. of nodes}  & $\mathbf{\langle k \rangle}$ & \textbf{Ref.} \\
\midrule
C. elegans herm. & 279 & 16.4  & \cite{github}\\
C. elegans male & 269 & 21.6  & \cite{Cook2019}\\
C. intestinalis 1 & 222 & 20.3  & \cite{Ryan2016}\\
C. intestinalis 2 & 138 &  6  & \cite{Ryan2016}\\
Drosophila medulla & 1770 & 10.1 & \cite{Takemura2013} \\
Drosophila con. mat. & 365  & 32.6 & \cite{Neurodata} \\
Drosophila adj. & 49 & 43 & \cite{Neurodata}  \\
Human DS01216 1 & 1121  & 32.4 & \cite{Neurodata}  \\
Human Talairach 1 & 959  & 44.7 & \cite{Neurodata}  \\
Human DS01216 2 & 1121  & 30.9 & \cite{Neurodata}  \\
Human Talairach 2 & 959  & 43.2 & \cite{Neurodata}  \\
Human DS01216 3 & 1121  & 32.2 & \cite{Neurodata}  \\
Human Talairach 3 & 959  & 45 & \cite{Neurodata}  \\
Human DS01216 4 & 1121  & 29.6 & \cite{Neurodata}  \\
Human Talairach 4 & 959  & 40.2 & \cite{Neurodata}  \\
Human DS01216 5 & 1121  & 30.1 & \cite{Neurodata}  \\
Human Talairach 5 & 959  & 42.5 & \cite{Neurodata}  \\
Human Rockland 1 & 188  & 13.2  & \cite{Hagmann2008}\\
Human Rockland 2 & 188  & 14.8  & \cite{Hagmann2008}\\
Human Rockland 3 & 188  & 14.4  & \cite{Hagmann2008}\\
Human Rockland 4 & 188  & 12.8  & \cite{Hagmann2008}\\
Human Rockland 5 & 188  & 16.7  & \cite{Hagmann2008}\\
Human Rockland 6 & 188  & 16.6  & \cite{Hagmann2008}\\
Human Rockland 7 & 188  & 16.0  & \cite{Hagmann2008}\\
Human Rockland 8 & 188  & 13.5  & \cite{Hagmann2008}\\
Human Rockland 9 & 188  & 15.5  & \cite{Hagmann2008}\\
Human Rockland 10 & 188  & 18.1  & \cite{Hagmann2008}\\
Human Hagmann 1 & 998  & 29.2  & \cite{Hagmann2008}\\
Human Hagmann 2 & 998  & 29.5  & \cite{Hagmann2008}\\
Human Hagmann 3 & 998  & 29.4  & \cite{Hagmann2008}\\
Human Hagmann 4 & 998  & 27.3  & \cite{Hagmann2008}\\
Human Hagmann 5 & 998  & 28.6  & \cite{Hagmann2008}\\
Mouse kasthuri & 987  & 3.1  & \cite{Neurodata}\\
Mouse brain & 213  & 151.1  & \cite{Neurodata}\\
Mouse retina & 1076  & 168.8  & \cite{github2}\\
Mouse visual cortex 1 & 29  & 3  & \cite{Bock2011}\\
Mouse visual cortex 2 & 193  & 2.2  & \cite{Bock2011}\\
Platynereis dumerilli & 2728 & 7.9 & \cite{Veraszto2020}\\
Rattus norvegicus 1 & 503 & 91.6  & \cite{Neurodata}\\
Rattus norvegicus 2 & 503  & 98.2  & \cite{Neurodata}\\
Rattus norvegicus 3 & 503  & 105.4  & \cite{Neurodata}\\
Rhesus brain 1 & 242  & 25.2  & \cite{Harriger2012}\\
Rhesus brain 2 & 91  & 12.8  & \cite{Neurodata}\\
Rhesus cerebral cortex & 91  & 30.8  & \cite{Neurodata}\\
Rhesus interareal cort. net. & 93  & 48.6  & \cite{Neurodata}\\
\bottomrule
\end{tabular}
\caption{List of connectomes of C. elegans, C. intestinalis, Drosophila, Human, Mouse, Platynereis dumerilly, Rattus norvegicus and Rhesus macaque. For each connectome the label of the network, the number of nodes of the network and the average degree $\langle k \rangle$ of the network are shown.}
\label{tab:info_connect}
\end{table}

\subsection*{Connectome-based RC}
RC is a Recurrent Neural Network passively excited by the input signal and that maintains, in its internal state, a non-linear transformation of the input history. The output signal is then generated as a linear combination of the neuron’s signals using a set of readout units that are trained to produce the desired target output. The key point is that the high-dimensional reservoir contains a representation of the input that is sufficiently rich to obtain the target output with a cheap linear learning procedure.
Let's consider a training data set of ordered input vectors $\vec{u}(n) \in \mathbb{R}^{N_u}$, $n=1,\dots,T$ each with its corresponding target vector $\vec{y}_{target}(n) \in \mathbb{R}^{N_y}$.
The goal is to learn the function
\begin{equation}
    \vec{y}(n) = y(\dots,\vec{u}(n-1),\vec{u}(n))
\end{equation}
that maps the input vector into the output vector minimizing an error function $E(\vec{y},\vec{y}_{target})$.

Firstly, the input vector is nonlinearly mapped into the reservoir, which is composed of $N$ neurons. The activation of each neuron determines the reservoir \textit{state vector} $\vec{x}(n) \in \mathbb{R}^{N}$. The input vector expansion can be written as:
\begin{equation}
    \vec{x}(n) = x(\dots,\vec{u}(n-1),\vec{u}(n)) = x(\vec{x}(n-1),\vec{u}(n))
\end{equation}
with the expansion function that has memory, namely it depends on previous input vectors. 
In this work, the following form for the expansion function is chosen
\begin{equation}
        \vec{x}(n) = (1-\alpha) \vec{x}(n-1) +
        \alpha \tanh ( W_{in} \vec{u'}(n) + 
        W \vec{x}(n-1) + \vec{b} )
        \label{eq:state_vec}
\end{equation}
where $\alpha$ is the leakage rate, $\vec{u'}(n) = (1, u_0 (n), \dots, u_{N_u} (n) )^T$ is the $n$-th biased input vector and  $\vec{b} \in \mathbb{R}^{N}$ is a bias.
Hyperbolic tangent function is used as neuron activation function. 
Secondly, the state vector is mapped into the output with a linear transformation
\begin{equation}
    \vec{y}(n) = W_{out}\vec{x}(n)
    \label{eq:train}
\end{equation}
where $W_{out} \in \mathbb{R}^{N_y \times N}$ is the output weights matrix. 
Learning procedure consists in finding the optimal values of the output weights matrix. The vector states are sequentially computed for all points in the training data set and are stored in the matrix $X = (\vec{x}(1), \dots, \vec{x}(T) )$. Given the target vectors $\vec{Y}_{target} = (\vec{y}_{target}(1), \dots, \vec{y}_{target}(T))$, the optimal output weights matrix can be computed as:
\begin{equation}
    W_{out} = \vec{Y}_{target} X^T (XX^T + \beta^2 \mathbb{I})^{-1}
    \label{eq:learning}
\end{equation}
where $\beta$ is a regularization parameter.
In this work, RC networks are connectomes provided as edge lists, namely lists of two indexes specifying the connected nodes. The weight of the connection is not considered.
Connectome data are pre-processed to obtain undirected unweighted connected graphs without multiple edges and self-loops. From each edge list the adjacency matrix $A_{conn}$ of the network is computed. 
Given a connectome with $N$ neurons described by the adjacency matrix $A_{conn} \in \mathbb{R}^{N \times N}$, the weight matrix $W$ of the reservoir is computed as element-wise multiplication
\begin{equation}
    W = A_{conn} W_{weight},
\end{equation}
where $W_{weight}$ is drawn from a uniform distribution in $[-1,1]$. Eventually, $W$ is rescaled as follows:
\begin{equation}
    W = W \frac{\rho_{exp}}{\rho(W)}
\end{equation}
with $\rho_{exp} = 1.25$ to set the value of the spectral radius to obtain the desired amount of memory and nonlinearity required.

The input matrix $W_{in}$ is  drawn from a uniform distribution in $\big[-\frac{1}{2},\frac{1}{2}\big]$. Moreover, the leakage rate is set to $\alpha = 0.3$ and the regularization parameter is set to be $\beta = 10^{-8}$. The bias vector $\vec{b}$ elements are sampled from a uniform distribution in $[-\frac{1}{2},\frac{1}{2}]$ and multiplied by a bias intensity factor of 0.1.

\subsection*{RC training and testing}
Each signal is a series $\vec{U} = (u(1),\dots,u(T))$ of one dimensional vectors $u(n) \in \mathbb{R}$. 
A transient period is of $L_{trans} = 100$ points of the signal is considered 
\begin{equation}
    \vec{U}_{trans} = (u(1),\dots,u(L_{trans}));
\end{equation}
the reservoir dynamics runs, as in Eq. \ref{eq:state_vec}, to initialize the activation pattern and no state vector is stored in this phase.

Then, the system runs with $L_{train} = 2500$ points of the signal 
\begin{equation}
    \vec{U}_{train} = (u(L_{trans}+1),\dots,u(L_{trans}+L_{train}))
\end{equation}
and activation states are stored in matrix $X$. 

The columns of $X$ may be correlated with each other so, in order to evaluate the number of uncorrelated vector states, we compute the rank of the covariance matrix of $X$ as 
\begin{equation}
    \Gamma= \text{rank}(X^TX)
\end{equation}

The output weight matrix $W_{out}$ can now be trained for the one-step-ahead prediction task as in Eq. \ref{eq:train}, using as target vector the input vector shifted forward of one unit 
\begin{equation}
    \vec{Y}_{target} = (u(L_{trans}+2),\dots,u(L_{trans}+L_{train}+1)).
\end{equation}

The trained RC is assessed by analyzing the performances on a test set of $L_{test} = 5000$ points of the signal
\begin{equation}
    \vec{U}_{test} = (u(L_{trans}+L_{train}+1),\dots,u(L_{trans}+L_{train}+L_{test})).
\end{equation}

The predictions are stored in the vector $\vec{Y}_{pred}$ and the performances are estimated based on the Mean Square Error (MSE) computed over a number $\tau$ of points as in the following equation:
 \begin{equation}
      MSE(\tau) = \frac{\sum_{i=1}^{\tau} (\vec{Y}_{pred,i} - \vec{Y}_{target,i})^2}{\tau};
      \label{eq:mse}
\end{equation}
and in particular the normalized MSE:
\begin{equation}
    MSE^* = \frac{MSE}{\text{std}(\vec{Y}_{target})}.
      \label{eq:mse_norm}
\end{equation}

For each connectome and input signal, the algorithm is run with 20 different random initializations of the weights and the average performances over all the realizations are computed.

Differences in performances between empirical connectomes and synthetic null models are evaluated using the Z-score as in the following equation:
\begin{equation}
    Z = \frac{\overline{MSE}_{syn} - \overline{MSE}_{emp}}{\sqrt{\sigma_{MSE_{syn}}^{2} + \sigma_{MSE_{emp}}^{2}}}
\end{equation}
where $\overline{MSE}$ is the the mean value of the normalized MSE over different realizations and $\sigma_{MSE}$ is the standard deviation.

\subsection*{Input signals}
Different time series (\textit{signals}) are used as inputs. Each signal is a series $\vec{U} = (u(1),\dots,u(T))$ of one dimensional vectors $u(n) \in \mathbb{R}$. 
Four signals are considered: two stochastic signals (white noise, fractional Brownian noise) and 2 deterministic chaotic signals (Ikeda and Mackey-Glass). 

The fractional Brownian motion is a generalization of the Brownian motion that consists in a continuous-time Gaussian process where the increments are not independent \cite{Mandelbrot1968}. The Brownian motion describes the random motion of particles suspended in a medium at thermal equilibrium, namely with no preferential direction of flow. The motion is described by the following equation:

\begin{equation}
\begin{split}
& B_H(t) = B_H (0)  \\
 + & \frac{1}{\Gamma(H+1/2)} \bigg\{ \int_{-\infty}^0 \left[ (t-s)^{H-1/2} - (-s)^{H-1/2} \right] dB(s) \\ 
+ & \int_0^t (t-s)^{H-1/2} dB(s) \bigg\} 
\end{split}   
\end{equation}

where the integration is with respect to the white noise measure $dB(s)$. $H$ is a parameter satisfying $0<H<1$; note that with $H=1/2$ the Brownian motion is recovered. 
The increment process can be written as
\begin{equation}
    X(t) = B_H (t+1) - B_H (t).
\end{equation}

Deterministic chaotic signals are generated by dynamical systems which appear to show a completely chaotic random behavior. Though, their behavior is deterministic because it can be fully determined by time evolution equations and initial conditions alone \cite{Lynch2018}. Chaotic systems are usually characterized by a long-term aperiodic behavior and a strong sensitivity to initial conditions \cite{Lynch2018}. The following systems are considered to generate the signals.
\begin{itemize}

\item Ikeda map is derived as a model of the light path across a non-linear optical resonator \cite{Ikeda1979}. The equations are:
\begin{equation}
    \begin{split}
    x_{n+1} & = A + B \left( x_n \cos\lvert E_n\rvert^2 - y_n \sin \lvert E_n\rvert^2 \right) \\
    y_{n+1} & = B \left( x_n \sin\lvert E_n\rvert^2 + y_n \cos \lvert E_n\rvert^2 \right)\\
    E_{n} & = x_n + i y_n 
    \end{split}   
\end{equation}
where $A$, $B$ are constants.
    
\item Mackey-Glass equations are delay differential equations designed to mimic both healthy and pathological behavior in some biological systems such as the variation in the relative quantity of mature cells in the blood \cite{Mackey1977}. The equation is:
    \begin{equation}
    \frac{dx}{dt} = \frac{\beta x(t-\tau)}{1+x(t-\tau)^n - \delta x(t)} 
    \end{equation}
with $\tau$ and $n$ constants and $x$ represents the density of cells over time.
\end{itemize}
white random noise is added to the signals to provide the RC input vectors with a predefined Signal to Noise Ratio (SNR) of 100. Moreover, each signal is normalized in the interval $\big[ -\frac{1}{2} , \frac{1}{2} \big]$

\end{small}


\bibliography{biblio}


\renewcommand*{\thefigure}{S\arabic{figure}}
\renewcommand*{\thetable}{S\arabic{table}}

\setcounter{figure}{0} 
\setcounter{table}{0} 

\begin{figure*}
\centering
\includegraphics[width=1\linewidth]{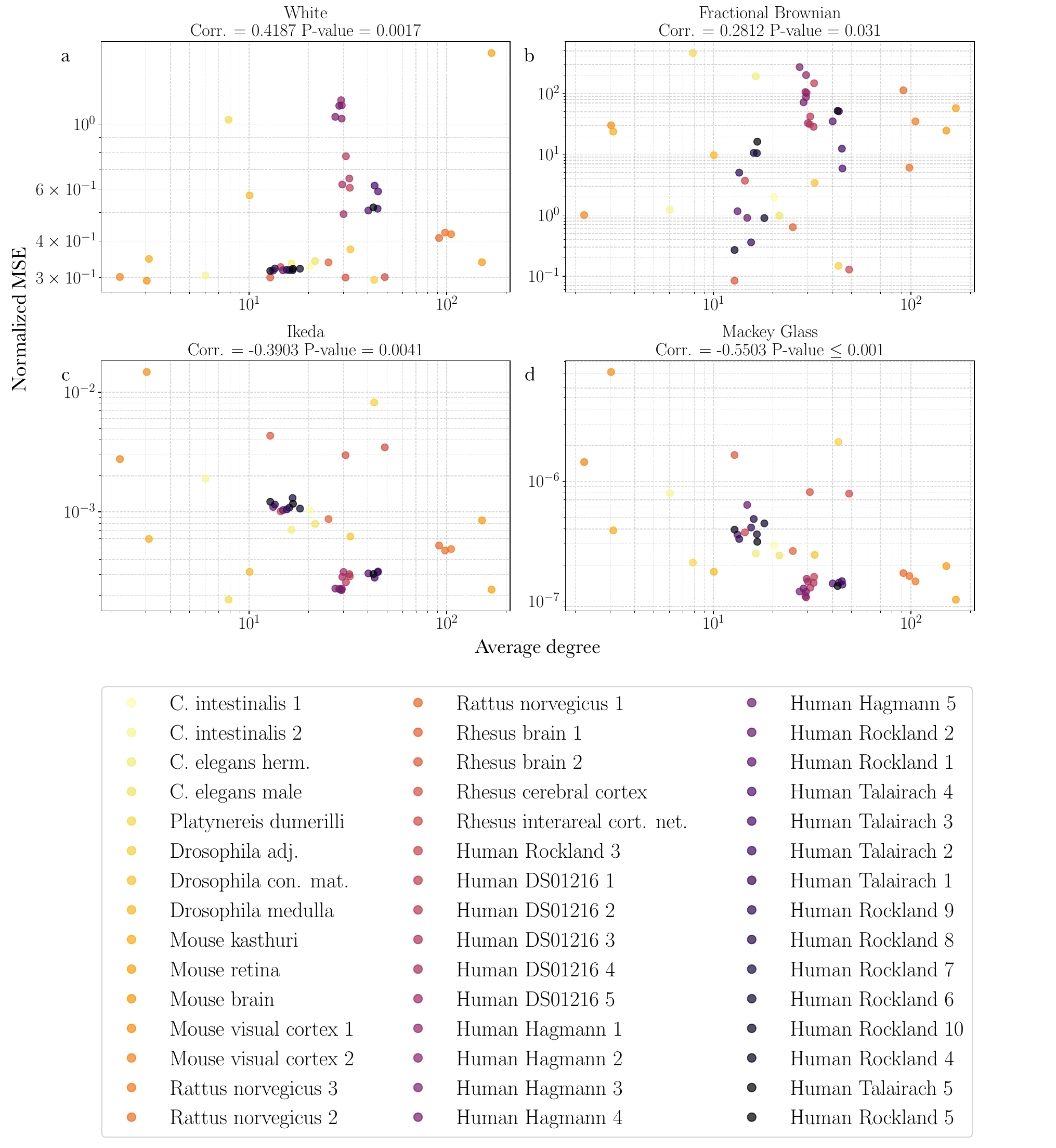}
\caption{\textbf{Correlations between normalized MSE and average degree.} a) White, b) Fractional Brownian, c) Ikeda and d) Mackey Glass. The color darkness of markers is proportional to the total number of neurons in the organisms brain. Correlations are significant ($p-value \leq 0.05$) for all signals. Positive values are observed for White and Fractional Brownian signals while negative values are observed for Ikeda and Mackey Glass signals.
}
\label{fig:figureS1}
\end{figure*}

\begin{figure*}
\centering
    \includegraphics[width=1\linewidth]{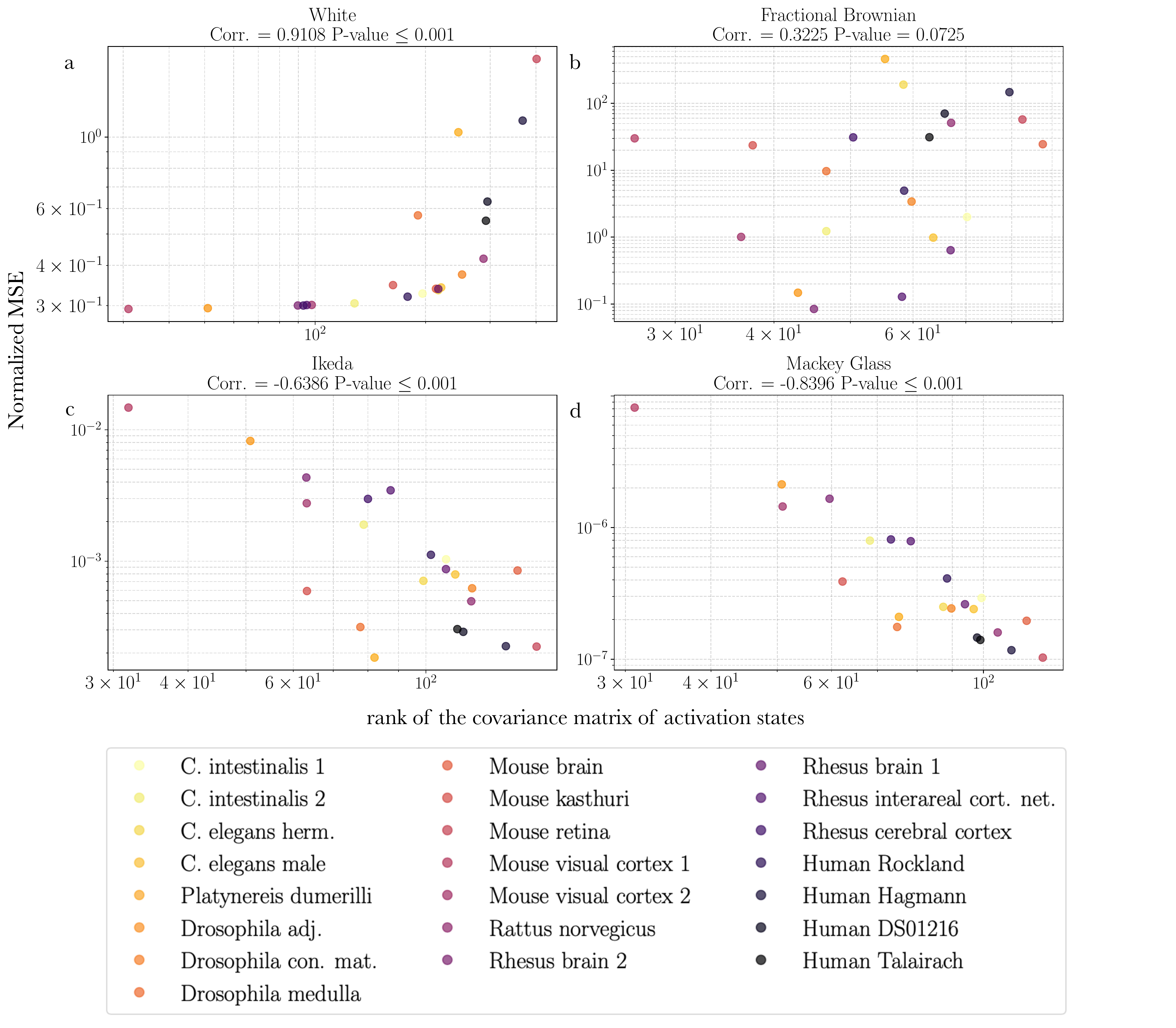}
    \caption{\textbf{Correlations between normalized MSE and rank of the covariance of matrix of activation states.}. Fa) White, b) Fractional Brownian, c) Ikeda and d) Mackey Glass. Each dot represents the average across all networks of the same organism with the same number of nodes.The color darkness of markers is proportional to the total number of neurons in the organisms brain. Correlations are found significant ($p-value \leq 0.05$) for all signals, except for the fractional Brownian signal. Positive values are observed for white and fractional Brownian signals while negative values are observed for Ikeda and Mackey Glass signals.
 \label{fig:figureS2}}
\end{figure*}

\begin{figure*}
\centering
    \includegraphics[width=1\linewidth]{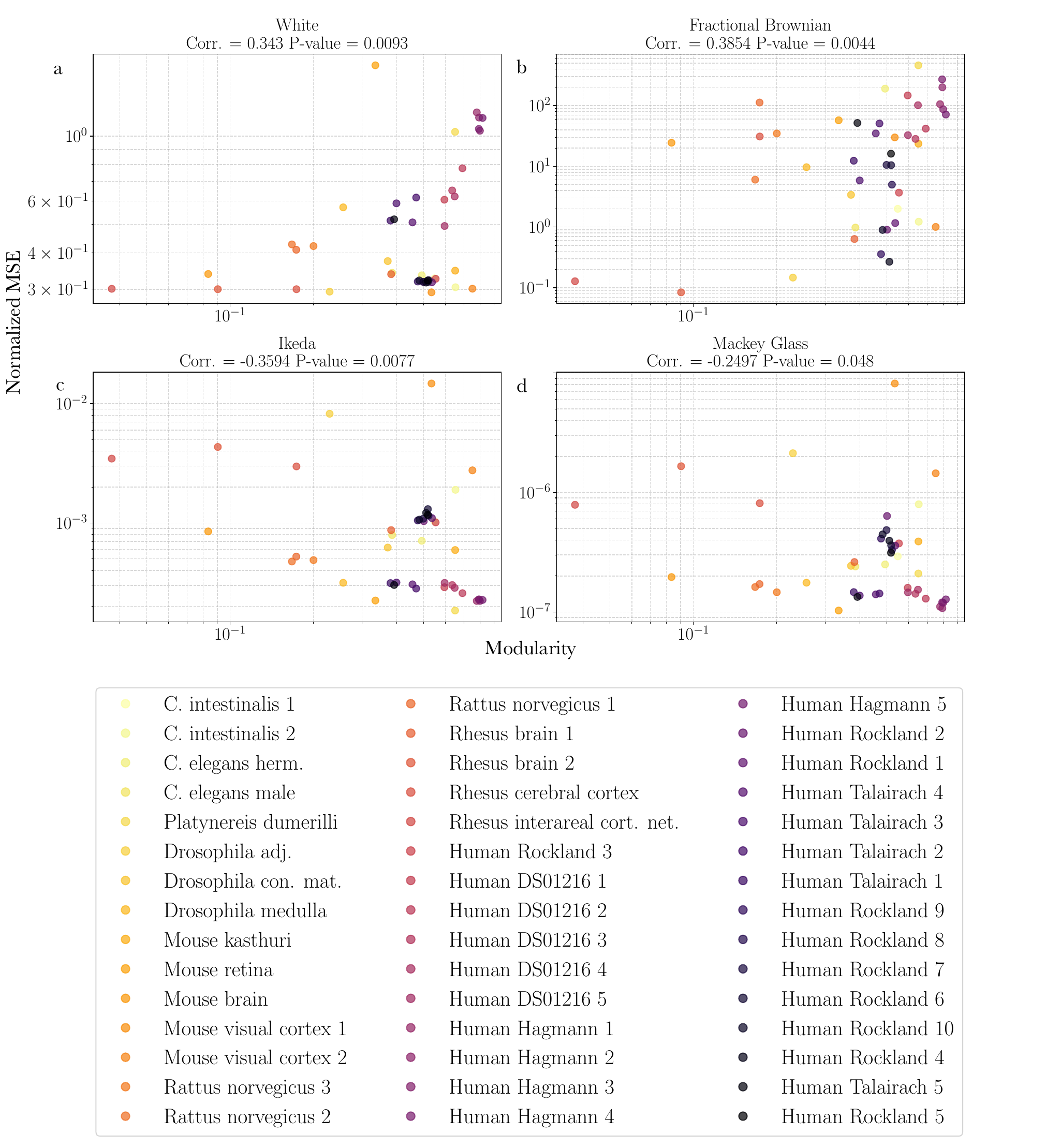}
    \caption{\textbf{Correlations between normalized MSE and modularity.} a) White, b) Fractional Brownian, c) Ikeda and d) Mackey Glass. The color darkness of markers is proportional to the total number of neurons in the organisms brain. Correlations are significant ($p-value \leq 0.05$) for all signals. Positive values are observed for white and fractional Brownian signals while negative values are observed for Ikeda and Mackey Glass signals.
 \label{fig:figureS3}}
\end{figure*}

\end{document}